# Fisher-information limits of detector-bandwidth-efficient 3D light-field microscopy

Liang Gao

Department of Bioengineering, University of California, Los Angeles, CA, 90095

**Abstract**

Light-field microscopy enables snapshot volumetric imaging, but its information rate is constrained by both optical encoding and detector readout architecture. Here we develop a task-dependent Fisher-information framework that evaluates optical encoders relative to the detector resource limiting acquisition throughput. We compare full Fourier light-field microscopy (FLFM), squeezed light-field microscopy (SLIM), and frame-rate-matched FLFM under a common optical geometry, photon budget, and row-limited camera model. Sparse scenes are analyzed using a 3D point-emitter Fisher matrix, and dense scenes using Fourier-mode information on tilted spectral slices. At $s = 0.25$, SLIM provides 2.40× higher axial Fisher information per camera bandwidth and 1.89× higher 3D D-optimal position information than frame-rate-matched FLFM. For dense scenes, it provides 1.85× higher integrated Fourier-mode Fisher information per bandwidth, 4× greater axial-frequency extent, and approximately 11× larger projected lateral hard-support area. Sweeps over compression factor and view tilt show that these advantages reflect a general detector-allocation principle rather than a specific operating point. More broadly, the framework can be adapted to other camera architectures by incorporating architecture-specific measurement models and detector-throughput costs, providing a general basis for co-designing optical encoding, scene statistics, and camera readout.

**Correspondence:** gaol@ucla.edu

## 1. Introduction

Light-field microscopy records multiple angular perspectives of a specimen in a single camera exposure, enabling volumetric reconstruction without mechanical axial scanning[1-3]. This snapshot acquisition is attractive for imaging rapid three-dimensional (3D) dynamics because volumetric rate can approach the native camera frame rate. The same parallelism, however, creates a detector-allocation trade-off: spatial and angular information must share a finite detection bandwidth. At high acquisition rates, detector readout can therefore become a primary constraint on both temporal resolution and recoverable spatial information.

Squeezed light-field microscopy (SLIM) addresses this allocation problem through disparity-aware anisotropic compression[4]. Each perspective view is first rotated so that its depth-dependent disparity is aligned with a common detector axis, and only the orthogonal direction is subsequently compressed. Native sampling is therefore retained along the disparity-sensitive coordinate while transverse sampling is reduced. Because the preserved axes are distributed among complementary view azimuths, the combined measurements retain broad 3D spatial-frequency coverage despite using fewer active detector rows. SLIM thus redistributes detector sampling according to the directional information content of the light field rather than reducing both image dimensions uniformly.

Evaluating such an encoder requires an information metric rather than a detector-occupancy metric alone. Fisher information (FI) is well suited to this purpose because it quantifies the sensitivity of the measured photon distribution to the parameters of interest and, through the Cramér-Rao lower bound (CRLB), sets the best attainable precision under a specified noise model[5-7]. FI is inherently task dependent. Sparse scenes are naturally parameterized by emitter coordinates; increasing source density introduces coupling among overlapping encoded point-spread functions (PSFs); and dense objects are more naturally described by voxel or spatial-frequency coefficients. Consequently, no single scene-independent scalar fully characterizes the information content of an optical encoder.

This task dependence is consistent with recent information-theoretic analyses showing that the preferred degree of optical multiplexing depends on object sparsity and detection noise[8]. We extend that perspective by arguing that the detector architecture must also enter the comparison. The relevant hardware resource is not necessarily total pixel count, but the detector degree of freedom that limits acquisition speed. This distinction is especially important for light-field microscopy, whose principal advantage is high-speed snapshot volumetric acquisition. In many column-parallel CMOS and sCMOS architectures, pixels are multiplexed row by row into parallel column readout circuits, such that frame time can depend predominantly on the number of active rows, whereas additional columns incur a substantially smaller readout penalty[4,9-11]. Equal pixel count therefore need not imply equal acquisition bandwidth.

This observation changes the appropriate comparison between SLIM and conventional Fourier light-field microscopy (FLFM). If SLIM compresses the row-limited detector dimension by a factor ($s$), producing an ($N \times sN$) image from each perspective view, the appropriate frame-rate-matched FLFM control is not an equal-pixel configuration. Instead, FLFM must be isotropically demagnified by the same factor ($s$), producing an ($sN \times sN$) view and therefore the same number of active rows and the same nominal frame rate as SLIM. SLIM deliberately retains the additional ($N$) samples along the column dimension because, under the assumed column-parallel readout architecture, those samples do not carry an equivalent camera-bandwidth cost. This motivates evaluating information efficiency in terms of FI per camera bandwidth, rather than FI per detector pixel.

Here we develop a task-dependent Fisher-information framework for comparing full FLFM, SLIM, and frame-rate-matched FLFM under a common three-view pupil geometry, detected photon budget, and camera readout model. Sparse scenes are analyzed with a full 3D point-emitter FIM after marginalizing emitter brightness, whereas dense scenes are analyzed in 3D Fourier space, where each perspective view contributes a tilted spectral slice. Task-relevant FI metrics are normalized by active-row count to quantify information per camera bandwidth. The resulting framework separates two design questions: the scene model determines which information is relevant, while the detector architecture determines the acquisition cost of obtaining it.

## 2. Fisher-information formulation

The framework developed here separates two coupled aspects of encoder performance. The scene model specifies the parameters to be estimated and therefore the appropriate Fisher-information representation, whereas the camera architecture specifies the detector resource that limits acquisition rate and should therefore be used for normalization. We first formulate FI at the measurement level, then define the detector-cost normalization used throughout the study, and finally specialize the analysis to sparse-emitter localization and dense spatial-frequency estimation.

### 2.1 Poisson measurement model and Fisher information

Let $y_i$ denote the detected photon count in camera pixel $i$, and let $\mu_i(\boldsymbol{\theta})$ denote its expected value for an unknown parameter vector $\boldsymbol{\theta}$. Under shot-noise-limited detection, the pixel measurements are modeled as independent Poisson random variables,

$$y_i \sim \text{Poisson}[\mu_i(\boldsymbol{\theta})]. \tag{1}$$

For statistically independent Poisson measurements, the Fisher information matrix (FIM) has elements

$$J_{mn}(\boldsymbol{\theta}) = \sum_i \frac{1}{\mu_i(\boldsymbol{\theta})}\left(\frac{\partial \mu_i}{\partial \theta_m}\right)\left(\frac{\partial \mu_i}{\partial \theta_n}\right). \tag{2}$$

Equation (2) can be interpreted as a noise-weighted sensitivity metric. For Poisson statistics, $\text{Var}(y_i) = \mu_i$, so the factor $1/\mu_i$ is the inverse measurement variance: a parameter-dependent change in a bright pixel is weighted against the larger shot noise carried by that pixel. Importantly, the derivative terms usually increase with photon number as well. If $\mu_i = N h_i(\boldsymbol{\theta})$, then $J \propto N$, so more detected photons still provide proportionally more FI.

A useful geometric interpretation follows by defining the noise-whitened sensitivity vector for parameter $\theta_m$ as

$$\mathbf{g}_m = \mathrm{diag}(\boldsymbol{\mu}^{-1/2})\frac{\partial \boldsymbol{\mu}}{\partial \theta_m}, \qquad J_{mn} = \mathbf{g}_m^{\mathrm{T}}\mathbf{g}_n.$$

The FIM is therefore the Gram matrix of the noise-whitened parameter-response vectors. Their norms quantify measurement sensitivity, while their relative orientations quantify parameter distinguishability. Nearly parallel response vectors indicate that two parameters perturb the measurements in similar ways and are therefore difficult to separate. Independent pixels and independent perspective views contribute additively. An encoder can thus increase useful information either by strengthening a parameter response or by providing complementary responses that reduce parameter degeneracy and increase the weaker eigenvalues of the FIM.

For any unbiased estimator $\widehat{\boldsymbol{\theta}}$, the Cramér–Rao inequality gives

$$\mathrm{Cov}(\widehat{\boldsymbol{\theta}}) \succcurlyeq \mathbf{J}^{-1},$$

and the corresponding lower bound for parameter $\theta_m$ is

$$\sigma_{\theta_m}^{\mathrm{CRLB}} = \sqrt{[\mathbf{J}^{-1}]_{mm}}.$$

The FIM is consequently not an intrinsic scalar property of an optical encoder; it depends on both the parameterization and the operating scene about which the likelihood is evaluated. Isolated emitters, overlapping sparse sources, and dense continuous objects are different inverse problems even when measured with identical optical hardware.

**2.2 Fisher information normalized by detector acquisition cost**

The efficiency of an optical encoder should be evaluated relative to the detector resource that limits acquisition throughput. Because this limiting resource depends on the camera architecture, we introduce a generic detector acquisition cost $C_{det}$, defined such that the frame time scales as

$$T_{\mathrm{frame}} \propto C_{det}.$$

The quantity $C_{det}$ may represent an active-row count, measured frame time, digitization load, data-transfer load, or another architecture-specific resource that determines acquisition rate.

Let $\Phi(\mathbf{J})$ denote a scalar Fisher-information measure appropriate to the task, such as effective axial FI, a D-optimal position metric, or integrated Fourier-mode FI. We define the detector-normalized information efficiency as

$$\eta_{det} = \frac{\Phi(\mathbf{J})}{C_{det}}. \qquad (3)$$

When $C_{det}$ is proportional to frame time, $\eta_{det}$ is proportional to the task-relevant FI acquired per unit time under a fixed per-frame photon budget. This formulation is deliberately architecture dependent: the detector cost should be chosen to represent the resource that actually limits measurement throughput.

For the column-parallel, row-limited camera model considered in this work, we approximate

$$T_{\mathrm{frame}} \approx \tau_{\mathrm{row}} N_{\mathrm{row}},$$

where $N_{\mathrm{row}}$ is the number of active detector rows and $\tau_{\mathrm{row}}$ is the effective readout time per row. Additional columns are assumed to be digitized largely in parallel and therefore incur a substantially smaller frame-time penalty. We consequently choose

$$C_{det} = N_{\mathrm{row}},$$

so that Eq. (3) reduces to

$$\eta_B = \frac{\Phi(\mathbf{J})}{N_{\mathrm{row}}},$$

which we refer to as *Fisher information per camera bandwidth* (FI/B). Here, camera bandwidth refers to the detector's pixel-digitization and data-transfer capacity—the rate at which measurements can be read out, converted, and transferred from the sensor—rather than to the temporal-frequency bandwidth of the imaged dynamics. Under the fixed per-frame photon budget used in this work, $\eta_B$ is proportional to the task-relevant FI that can be acquired per unit time when frame rate is limited by active-row readout. If photon flux or phototoxicity rather than per-frame photon number is held fixed, the photon budget must be rescaled with exposure time before interpreting FI/B as an absolute information rate.

This distinction is central to the comparison. Equal total pixel count does not imply equal frame rate for a row-limited detector. SLIM can retain additional column samples without incurring the same readout penalty as additional rows, whereas frame-rate-matched FLFM must reduce both image dimensions to achieve the same $C_{det} = N_{\mathrm{row}}$.

**2.3 Sparse-emitter Fisher information**

For a sparse fluorescent scene containing $K$ emitters, the expected count in pixel $i$ is written as

$$\mu_i = \sum_{k=1}^{K} a_k\, h_i(\mathbf{r}_k) + b_i,$$

where $\mathbf{r}_k = (x_k, y_k, z_k)$ is the 3D position of emitter $k$, $a_k$ is its expected detected signal, $h_i(\mathbf{r}_k)$ is the encoder-dependent fraction of that signal falling in pixel $i$, and $b_i$ is the expected background. The unknown parameter vector contains both position and brightness,

$$\boldsymbol{\theta} = (x_1, y_1, z_1, \dots, x_K, y_K, z_K, a_1, \dots, a_K)^{\mathrm{T}}.$$

The position derivatives in Eq. (2) are determined by the spatial derivatives of the encoded point-spread function; for example,

$$\frac{\partial \mu_i}{\partial x_k} = a_k \frac{\partial h_i}{\partial x_k}, \qquad \frac{\partial \mu_i}{\partial z_k} = a_k \frac{\partial h_i}{\partial z_k}, \qquad \frac{\partial \mu_i}{\partial a_k} = h_i.$$

Emitter brightness is not a parameter of interest in the present analysis and is therefore treated as a nuisance parameter. Partitioning the full FIM into position and brightness blocks,

$$\mathbf{J} = \begin{pmatrix} \mathbf{J}_{pp} & \mathbf{J}_{pa} \\ \mathbf{J}_{ap} & \mathbf{J}_{aa} \end{pmatrix},$$

the Fisher information available for position estimation after accounting for unknown brightness is given by the Schur complement

$$\mathbf{J}_{\mathrm{pos}} = \mathbf{J}_{pp} - \mathbf{J}_{pa}\mathbf{J}_{aa}^{-1}\mathbf{J}_{ap},$$

A pseudoinverse is used when required numerically. The subtraction has a direct interpretation: a measurement variation that can be explained by a change in emitter brightness cannot be counted independently as position information.

For $K = 1$, $\mathbf{J}_{\mathrm{pos}}$ is a $3 \times 3$ matrix; for $K > 1$, it is a $3K \times 3K$ matrix. Off-diagonal blocks couple different emitters, while off-diagonal elements within each block couple coordinates such as $x$ and $z$. These correlations are precisely the parameter degeneracies described in Sec. 2.1. As PSFs overlap, the response vectors of different emitters become more similar, the FIM becomes more poorly conditioned, and joint source localization becomes more difficult.

For a specific coordinate such as $z_k$, we report the effective FI after all other position coordinates have been marginalized,

$$J_{z_k,\mathrm{eff}} = \frac{1}{\left[\mathbf{J}_{\mathrm{pos}}^{-1}\right]_{z_k z_k}}, \qquad \sigma_{z_k}^{\mathrm{CRLB}} = \frac{1}{\sqrt{J_{z_k,\mathrm{eff}}}}.$$

This effective information is generally smaller than the raw diagonal element $J_{z_k z_k}$ whenever $z_k$ is correlated with other unknown coordinates.

To summarize balanced 3D localization performance, we additionally use the D-optimal metric

$$\Phi_D = \det(\mathbf{J}_{\mathrm{pos}})^{1/(3K)}, \qquad \eta_{B,D} = \frac{\Phi_D}{N_{\mathrm{row}}}.$$

Because the determinant is the product of the FIM eigenvalues, $\Phi_D$ is their geometric mean and penalizes a design that provides very high information along some parameter combinations but leaves others poorly constrained. We also monitor the smallest eigenvalue and condition number as measures of worst-mode information and degeneracy. This coordinate-based formulation is appropriate for isolated and crowded point sources, but it becomes unnatural for a dense continuous volume, for which individual emitter identities are no longer the relevant unknowns.

### 2.4 Fisher information for dense spatial-frequency modes

For a dense object, individual emitter identities are no longer the natural unknowns. We therefore parameterize the scene by its spatial-frequency content and consider a weak Fourier perturbation of a locally uniform fluorescent object,

$$\rho(\mathbf{r}) = \rho_0 + \epsilon\cos(\mathbf{k}\cdot\mathbf{r} + \psi),$$

where $\epsilon$ is the perturbation amplitude, $\psi$ is its phase, and $\mathbf{k} = (k_x, k_y, k_z)$ is the 3D spatial frequency. For a linear imaging system, the modulation transferred to view $j$ is proportional to the view transfer function $H_j(\mathbf{k})$. Equivalently, treating the two Fourier quadratures as real parameters, the noise-whitened response amplitude scales with $|H_j(\mathbf{k})|$.

For a weak perturbation on a locally uniform Poisson background, substitution into Eq. (2) gives a Fourier-mode FI that is proportional to the photon allocation in each view and to the squared transfer magnitude,

$$J_{\epsilon\epsilon}(\mathbf{k}) \propto \sum_j N_j \left|H_j(\mathbf{k})\right|^2.$$

With three views receiving equal fractions of a fixed total photon budget, $N_j = N_{\mathrm{ph}}/3$, common photon-dependent factors cancel in relative encoder comparisons. We therefore define the normalized Fourier-mode Fisher-information density as

$$I(\mathbf{k}) \propto \frac{1}{3}\sum_{j=1}^{3}\left|H_j(\mathbf{k})\right|^2. \tag{4}$$

Equation (4) is the dense-scene analogue of the additive FIM in Sec. 2.1. A frequency not transferred by any view has zero information, whereas a frequency transferred by more than one view can carry increased FI because statistically independent measurements add. Importantly, overlap and coverage are not equivalent: multi-view overlap increases the FI magnitude at an already sampled frequency, whereas extending the support of $H_j$ makes previously inaccessible object frequencies observable. The system-specific 3D support of $H_j(\mathbf{k})$ for full FLFM, SLIM, and frame-rate-matched FLFM is introduced in Sec. 3.3.

### 2.5 Integrated information, support bandwidth, and information efficiency

The frequency-resolved map $I(\mathbf{k})$ contains more information than any single scalar, so we report complementary quantities that distinguish *how much* information is acquired from *where* that information lies in 3D Fourier space.

The integrated Fourier-mode FI over the sampled slice surfaces $\mathcal{S}$ is

$$I_{\mathrm{int}} = \int_{\mathcal{S}} I\,(\mathbf{k})\,dA,$$

with camera-bandwidth-normalized form

$$\eta_{B,\mathrm{int}} = \frac{I_{\mathrm{int}}}{N_{\mathrm{row}}}.$$

This quantity measures the total Fourier-mode information collected per unit row-readout cost, but it is not itself a resolution metric. Because the incoherent optical transfer function decreases toward its cutoff, low and intermediate spatial frequencies can dominate the integral even when high-frequency support is limited.

We therefore also report the binary hard support

$$S_{\mathrm{hard}}(\mathbf{k}) = 1 \quad \text{if} \quad |H_j(\mathbf{k})| > 0 \text{ for at least one view } j,$$

and characterize its 3D slice area, projected lateral area, directional radial extent, and maximum axial frequency. In parallel, an FI-defined bandwidth is obtained by applying a common threshold to $I(\mathbf{k})$. The hard-support metrics answer whether a spatial frequency is sampled at all, whereas the FI-defined metrics additionally account for transfer strength and multi-view redundancy.

These distinctions are particularly important for anisotropic encoding. SLIM is not intended simply to maximize integrated FI; it reallocates detector bandwidth so that high spatial frequencies along the disparity-sensitive coordinate are retained while transverse sampling is reduced. We therefore report FI magnitude, FI per camera bandwidth, hard-support extent, directional isotropy, and multi-view redundancy as complementary descriptors rather than collapsing encoder performance into a single scalar.

## 3. Comparison design and simulation models

We compare full FLFM, SLIM, and frame-rate-matched FLFM under a common optical configuration, photon budget, and detector-bandwidth constraint. The comparison proceeds in two stages. First, the three encoders are placed on the same three-view acquisition geometry and the compressed configurations are matched by active-row count, the limiting resource in the assumed camera model. Second, we apply scene models that mirror the parameterizations of Secs. 2.3 and 2.4: a point-emitter model for sparse scenes and a 3D Fourier-slice model for dense scenes. Holding the pupil views and detected photons per frame fixed isolates the effect of detector mapping from illumination and photon-budget differences. The baseline simulation parameters are summarized in Supplementary Note 1.

### 3.1 Light-field encoders

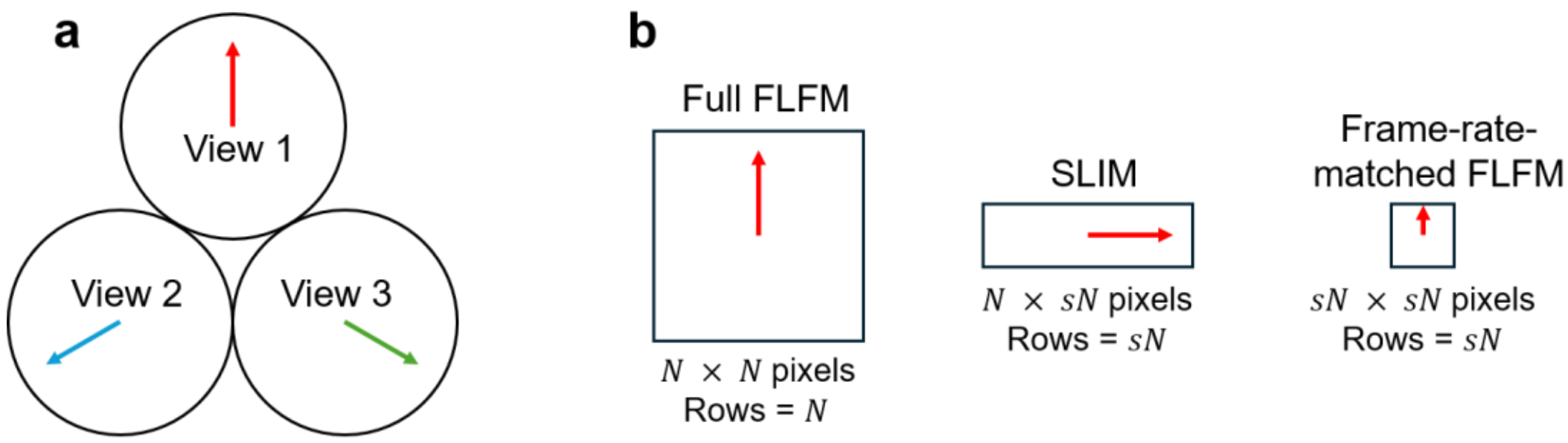

**Figure 1. Comparison of light-field encoders. (a)** Three-view pupil configuration. The arrow in each view denotes the corresponding depth-disparity direction. **(b)** Detector sampling for full FLFM, SLIM, and frame-rate-matched FLFM under a common three-view geometry and photon budget. Full FLFM records $(N \times N)$ samples per view. SLIM preserves $(N)$ samples along the disparity-aligned column axis while compressing the orthogonal row dimension to $(sN)$. Frame-rate-matched FLFM instead isotropically demagnifies each view to $(sN \times sN)$. Under the assumed column-parallel, row-limited readout architecture, SLIM and isotropically demagnified FLFM use the same number of active rows and therefore have the same nominal frame rate; their different column counts reflect distinct allocations of the non-rate-limiting detector dimension.

All three systems use the same pupil samples at azimuths φ = 0°, 120°, and 240° (Fig. 1(a)) and receive the same total detected signal photons per frame; only the mapping from each perspective view to the camera is changed (Fig. 1(b)). Full FLFM records an $N \times N$ image for every view. SLIM first rotates each view so that its depth-disparity direction is aligned with the detector-column axis and then squeezes only the orthogonal row direction by a factor $s$, producing $N \times sN$ samples per view. Frame-rate-matched FLFM applies no view-dependent rotation and instead demagnifies both image axes by $s$, producing $sN \times sN$ samples per view. Under the row-limited camera model of Sec. 2.2, the two compressed encoders therefore use the same active-row count and have the same nominal readout time. Equation 5 summarizes the resulting row counts, frame-rate scaling, and FI/B normalization.

$$N_{\text{row,full}} = N, \qquad N_{\text{row,SLIM}} = N_{\text{row,BW}} = sN, \qquad f_{\max} \propto N_{\text{row}}^{-1}, \qquad \eta_B = \frac{\Phi(\mathbf{J})}{N_{\text{row}}}. \tag{5}$$

For the baseline comparison, $N = 64$ and $s = 0.25$. Full FLFM therefore uses 64 active rows per view and 12,288 pixels across the three views. SLIM uses 16 rows per view while retaining 64 columns, for 3,072 total pixels. Frame-rate-matched FLFM also uses 16 rows per view but retains only 16 columns, for 768 total pixels. Thus SLIM and frame-rate-matched FLFM have the same nominal 4× frame-rate advantage over full FLFM in the assumed readout model, even though SLIM records four times as many pixels as the isotropic control. This difference is intentional: the comparison asks whether the non-limiting column dimension can be used to preserve task-relevant information without increasing the row-readout cost.

### 3.2 Sparse-scene point-emitter model

For the sparse-scene analysis, each native sub-view is represented by a pixel-integrated Gaussian PSF with σ = 0.90 full-FLFM camera pixels (FWHM ≈ 2.12 pixels). Axial displacement produces a view-dependent lateral disparity of 2.5 native camera pixels per normalized axial unit. Each emitter contributes 1,500 detected signal photons, and the total background is 2% of the signal. Identical emitter coordinates, photon budgets, and backgrounds are used for all three encoders so that differences in the FIM arise only from view rotation, detector sampling, and pixelation.

For isolated emitters, $z$ is swept from −1 to +1. At $s$ = 0.25, isotropic demagnification makes the frame-rate-matched FLFM PSF strongly undersampled, so a single centered emitter can exhibit an artificial dependence on detector subpixel phase. We therefore average all FIM-derived scalar metrics at each $z$ over a 4 × 4 grid of lateral offsets spanning one complete detector-pixel phase period of the frame-rate-matched output. In native full-FLFM coordinates, the offsets are $x, y$ = −1.5, −0.5, 0.5, 1.5; the same physical offsets are used for every encoder. Figure 2(a) reports the phase mean and the 10th–90th percentile range. To probe crowding, $K$ = 1, 2, 4, 8, and 16 emitters are then placed uniformly in a fixed 10 × 10 × 2 normalized volume, with 20 common random realizations at each $K$. These random lateral coordinates naturally sample detector phase while progressively increasing PSF overlap and parameter coupling.

The Gaussian model is intentionally mechanistic rather than a complete wave-optical description of the segmented pupil. Its purpose is to isolate the influence of disparity-aware rotation, anisotropic detector sampling, pixelation, and Poisson noise. Dense-object performance is therefore evaluated separately with the Fourier-domain model below rather than inferred from point-localization behavior.

### 3.3 Dense-scene 3D Fourier-slice model

To analyze dense scenes with the frequency-domain metric of Sec. 2.4, each perspective view is modeled as a tilted slice through the 3D object spectrum. Let α denote the polar view tilt and $\varphi_j$ the azimuth of view $j$. The view normal $\mathbf{n}_j$ is given by

$$\mathbf{n}_j = \left(\sin\alpha\cos\phi_j\,,\ \sin\alpha\sin\phi_j\,,\ \cos\alpha\right),$$

so the corresponding ideal Fourier-slice plane satisfies

$$\mathbf{k} \cdot \mathbf{n}_j = 0.$$

Within each slice, we use an orthonormal in-plane basis that separates the disparity-aligned coordinate $k_u$ from the transverse coordinate $k_v$. Their basis vectors and corresponding in-plane coordinates are given by Eqs. 6 and 7. This choice makes the detector mappings directly comparable because SLIM preserves sampling along $k_u$ while compressing $k_v$.

$$\mathbf{e}_{u,j} = (\cos\alpha\cos\phi_j\,,\ \cos\alpha\sin\phi_j\,,\ -\sin\alpha), \qquad \mathbf{e}_{v,j} = (-\sin\phi_j\,,\ \cos\phi_j\,,\ 0) \tag{6}$$

$$k_u = \mathbf{k} \cdot \mathbf{e}_{u,j}, \qquad k_v = \mathbf{k} \cdot \mathbf{e}_{v,j} \tag{7}$$

The native full-FLFM cutoff is normalized to $k_N$= 1. Full FLFM retains the complete unit disk on every slice. SLIM retains the native extent $|k_u| \leq 1$ along the disparity-sensitive direction while restricting the transverse extent to $|k_v| \leq$ s, with the support clipped by the native circular passband. Frame-rate-matched FLFM reduces both coordinates isotropically, giving ${k_u}^2 + {k_v}^2 \leq s^2$. A circular incoherent OTF envelope weights the transferred modes within these supports. Quantitative metrics are evaluated on zero-thickness slices; a narrow finite thickness is introduced only for 3D visualization and does not enter the reported slice-area or integrated-FI values.

Consistent with Sec. 2.5, we keep coverage and information magnitude separate. The native in-slice support describes the frequencies transmitted by a single view in its own Fourier plane. The projected hard-support map then asks whether a lateral frequency $(k_x, k_y)$ is sampled by at least one of the three tilted slices. Because a tilted circular disk projects to an ellipse, the three-view projection of full FLFM can show a weak sixfold boundary modulation even though each native view is exactly circular. Finally, the projected FI density weights the sampled frequencies by transfer magnitude and by the number of statistically independent views that contribute. These three quantities are used together in Sec. 4 to distinguish spatial-frequency reach from multi-view redundancy.

## 4. Results

We first examine the sparse limit, where FI directly quantifies sensitivity to emitter coordinates, and then increase source density to expose parameter coupling caused by PSF overlap. We subsequently consider dense objects, for which 3D spatial-frequency support and Fourier-mode FI are the appropriate observables. This progression addresses three distinct questions: how much local coordinate information is preserved, how that information degrades under source ambiguity, and where the retained information lies in 3D Fourier space.

### 4.1 Isolated emitters: axial information at matched camera bandwidth

In the isolated-emitter limit, SLIM retains essentially all the axial information available to full FLFM: the median effective axial FI is 10,473 for SLIM and 10,478 for full FLFM. Its total x- and y-information are each approximately 73% of the full-FLFM values, reflecting the transverse sampling sacrificed by one-axis squeezing. The ranking changes when readout cost is included. Because SLIM uses 16 active rows instead of 64, its axial FI/B is 4.00× that of full FLFM and its 3D D-optimal position FI/B is 3.24× higher. Relative to isotropically demagnified FLFM at the same nominal frame rate, SLIM provides 2.40× higher axial FI/B and 1.89× higher D-optimal FI/B. Table 1 summarizes the key quantitative metrics derived from this simulation.

As shown in Fig. 2(a), subpixel-phase averaging suppresses the strong grid-dependent oscillation of the severely undersampled frame-rate-matched PSF; the residual axial-FI variation over z is approximately 15% peak to trough. SLIM and full FLFM remain nearly depth independent because the disparity-sensitive coordinate remains approximately Nyquist sampled. Thus, SLIM trades some transverse information for preservation of the depth-sensitive response at substantially lower row-readout cost.

| Method | Rows | Pixels | $FI_x/B$ | $FI_y/B$ | $FI_z/B$ | CRLB $\sigma_x$ | CRLB $\sigma_y$ | CRLB $\sigma_z$ | 3D D-opt FI/B |
|---|---|---|---|---|---|---|---|---|---|
| FLFM full | 64 | 12,288 | 26.20 | 26.20 | 163.72 | 0.0244 | 0.0244 | 0.0098 | 48.25 |
| SLIM | 16 | 3,072 | 76.94 | 76.00 | 654.57 | 0.0286 | 0.0288 | 0.0098 | 156.4 |
| FLFM frame-rate-matched | 16 | 768 | 44.95 | 45.63 | 273.03 | 0.0388 | 0.0379 | 0.0155 | 82.89 |

**Table 1. Isolated-emitter localization metrics under the row-limited camera model.** Values are medians over z = −1 to +1 after averaging the FIM-derived scalar metrics at each z over the 4 × 4 lateral detector-phase ensemble described in Sec. 3.2. FI/B denotes FI divided by active-row count and is proportional to information rate under the assumed row-limited readout.

### 4.2 Crowded sparse emitters: information loss from parameter coupling

The isolated-emitter result measures local parameter sensitivity in the absence of source ambiguity. As the number of emitters increases, the encoded PSFs overlap and the off-diagonal blocks of the joint position FIM grow. The D-optimal FI/B therefore decreases for all three encoders as source density increases, while the condition number rises, consistent with increasing parameter degeneracy (Fig. 2(b)). SLIM retains substantially more joint position information than frame-rate-matched FLFM across the tested densities because the isotropic control discards both image dimensions to reach the same active-row count. At the highest density, the frame-rate-matched FIM becomes particularly ill conditioned. This regime is nevertheless a source-association problem, namely joint estimation of multiple continuous emitter coordinates, rather than a dense-volume resolution metric.

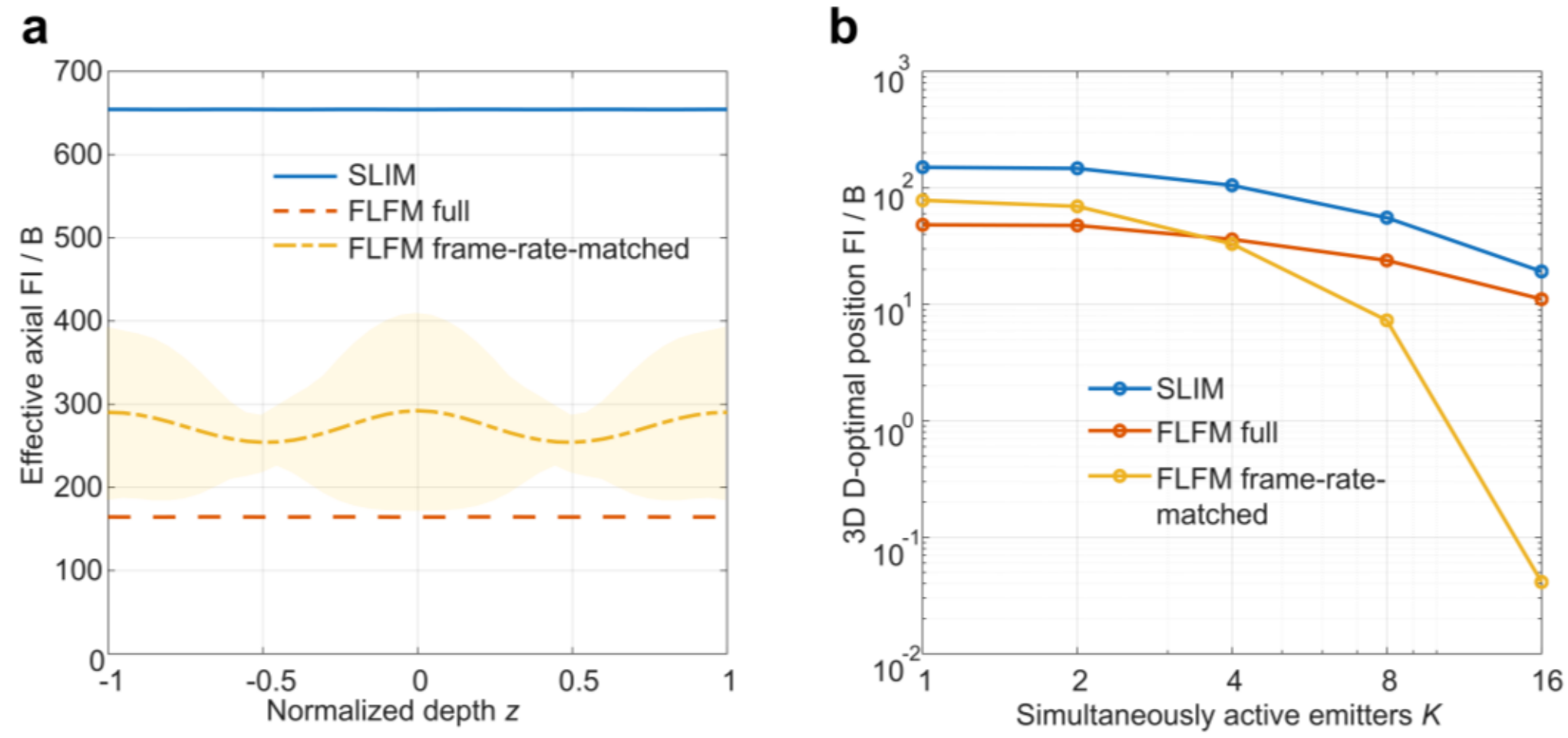

**Figure 2. Sparse-scene Fisher information normalized by camera row bandwidth.** (a) Effective axial FI/B for an isolated emitter versus normalized depth. Solid curves show the mean over 16 lateral detector phases and shaded bands show the 10th–90th percentile range. Phase averaging suppresses the grid-dependent oscillation of the strongly undersampled frame-rate-matched control. SLIM preserves nearly the full-FLFM axial information while using one quarter as many active rows and remains substantially above isotropically demagnified FLFM at the same nominal frame rate. (b) 3D D-optimal position FI/B versus the number of simultaneously active emitters. Increasing overlap reduces joint coordinate separability for all methods; the random lateral positions naturally average over detector phase.

### 4.3 Dense Fourier modes: preservation of 3D spatial bandwidth

The parameterization changes when the scene is treated as a continuous object rather than as identifiable emitters. As shown in Fig. 3, each perspective view now contributes a differently tilted slice through the 3D object spectrum. Full FLFM retains a complete in-plane disk on each slice. SLIM preserves the full native extent along $k_u$ but compresses the transverse coordinate to $s$, whereas frame-rate-matched FLFM reduces both in-plane coordinates to s in order to use the same active-row count. Because $k_u$ is the in-plane coordinate coupled to $k_z$ by the slice tilt, this difference predicts that SLIM should preserve substantially greater axial-frequency reach at the same nominal camera bandwidth.

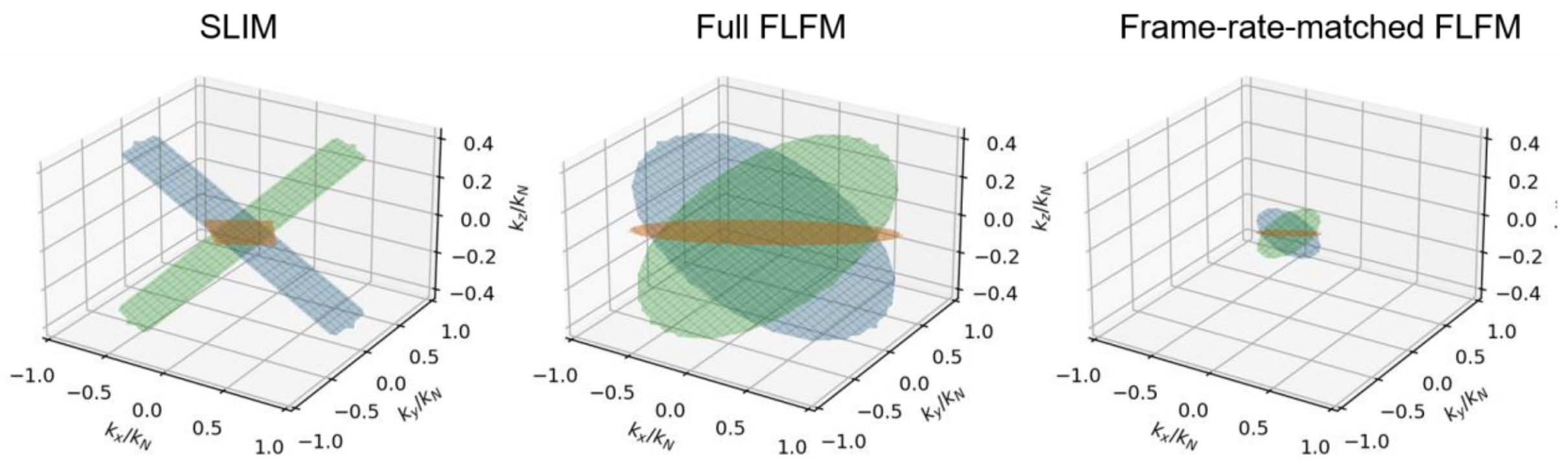

**Figure 3. 3D Fourier-slice hard-support geometry at equal photon budget (illustrative α = 20°).** Full FLFM retains a unit disk on every tilted slice. SLIM retains an elongated region with native $k_u$ reach and transverse extent $s$, while frame-rate-matched FLFM retains an isotropically reduced disk of radius $s$. SLIM and frame-rate-matched FLFM use the same active-row count; their different supports therefore show how anisotropic versus isotropic detector allocation redistributes 3D spatial bandwidth at the same nominal frame rate.

| Method | Rows | Rel. frame rate | Hard slice area | FI area ≥0.01 | Integrated FI | Integrated FI/B | max $\lvert k_z\rvert/k_N$ |
|---|---|---|---|---|---|---|---|
| FLFM full | 64 | 1× | 9.424 | 6.113 | 0.361 | 0.00564 | 0.342 |
| SLIM | 16 | 4× | 2.977 | 2.383 | 0.229 | 0.01431 | 0.342 |
| FLFM frame-rate-matched | 16 | 4× | 0.589 | 0.589 | 0.123 | 0.00771 | 0.0855 |

**Table 2. Zero-thickness 3D Fourier-slice metrics for α = 20° under the row-limited camera model.** Hard-support and FI areas are evaluated on the tilted slice surfaces. Integrated FI/B is the surface-integrated Fourier-mode FI divided by active-row count. The FI-area threshold is 0.01 relative to full-FLFM DC FI.

The hard-support metrics confirm this geometric advantage (Table 2). At $s$ = 0.25, SLIM samples 2.977 normalized units of three-slice surface area, compared with 0.589 for frame-rate-matched FLFM, a 5.05× increase. Full FLFM remains larger at 9.424 because it retains the complete unit disk in every view. More important for depth encoding, SLIM and full FLFM reach the same hard axial-frequency extent, $|k_z|_{\max} = 0.342k_N$, whereas frame-rate-matched FLFM reaches only $0.0855k_N$. SLIM therefore preserves a 4× larger axial-frequency range than the isotropic control at the same active-row count and nominal frame rate.

The FI-weighted comparison leads to the same qualitative conclusion while emphasizing information efficiency rather than maximum spectral reach. Full FLFM has the largest integrated FI per frame (0.361), followed by SLIM (0.229) and frame-rate-matched FLFM (0.123). After normalization by active rows, the corresponding FI/B values are 0.00564, 0.0143, and 0.00771. SLIM therefore provides 1.85× higher integrated FI/B than frame-rate-matched FLFM and 2.54× that of full FLFM. This distinction is essential: integrated FI/B quantifies information collected per row-readout cost, whereas hard-support extent identifies the spatial frequencies that remain accessible at all.

### 4.4 Projected support separates coverage from Fisher-information redundancy

Lateral projections provide an intuitive view of the 3D supports, but coverage and redundancy must remain distinct. The projected hard support (Fig. 4(a)) is a binary union: a lateral frequency is included if at least one tilted slice samples it. The projected FI density (Fig. 4(b)) instead weights that frequency by transfer magnitude and by the number of independent views that contribute. Full FLFM therefore retains an approximately full native lateral support despite the bright sixfold structure in its FI-density projection; those arms arise from overlap among tilted slices rather than from isolated passbands. Frame-rate-matched FLFM contracts to a small isotropic region, whereas SLIM forms a broader complementary union from the three anisotropic supports.

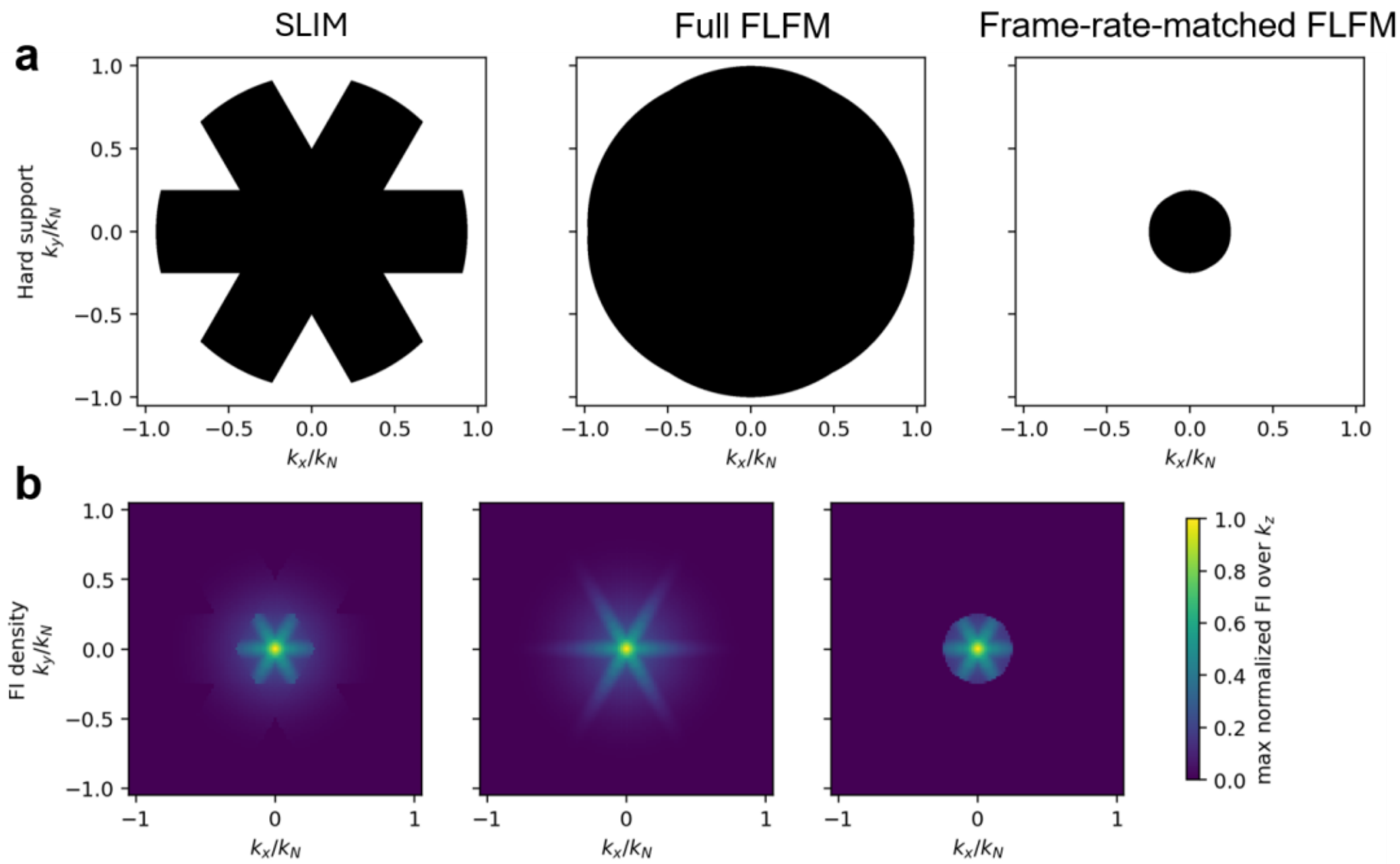

**Figure 4. Projected hard support and projected FI density under the camera-bandwidth comparison.** (a) binary $k_x - k_y$ union of frequencies sampled by at least one of the three tilted Fourier slices. (b) maximum-over-$k_z$ normalized FI density. Full FLFM remains broadly supported because each native view is unsqueezed; frame-rate-matched FLFM contracts isotropically; and SLIM forms a complementary anisotropic union. Bright radial structure in the FI-density maps marks multi-view redundancy and should not be interpreted as missing support between the arms.

| Method | Projected hard area | Hard $k_{min}$ | Hard $k_{max}$ | FI $k_{min}$ | FI $k_{mean}$ | FI $k_{max}$ |
|---|---|---|---|---|---|---|
| FLFM full | 3.106 | 0.983 | 1.000 | 0.705 | 0.718 | 0.764 |
| SLIM | 2.142 | 0.502 | 0.943 | 0.502 | 0.646 | 0.677 |
| FLFM frame-rate-matched | 0.194 | 0.245 | 0.250 | 0.245 | 0.248 | 0.250 |

**Table 3. Lateral projected-support and FI-defined bandwidth metrics.** Hard $k_{min}$ and $k_{max}$ denote the minimum and maximum radial boundaries of the binary three-view projected support. FI $k_{min}$, $k_{mean}$, and $k_{max}$ are obtained using a common threshold of 0.01 relative to full-FLFM DC FI. Frame-rate-matched FLFM is isotropically limited to radius $s = 0.25$; the small angular variation of full FLFM arises only from projection of tilted circular disks.

The projected metrics make the redistribution of bandwidth explicit. In a single view's own Fourier plane, full FLFM is exactly circular with radius $k_N$ and frame-rate-matched FLFM is exactly circular with radius $sk_N$ (Fig. 5(a)). After the three tilted views are projected onto $k_x - k_y$ plane, full FLFM remains nearly circular, whereas SLIM forms a complementary sixfold union (Fig. 5(b)). At α = 20°, the projected hard-support area is 2.142 $k_N{}^2$ for SLIM and 0.194 $k_N{}^2$ for frame-rate-matched FLFM, an 11.0× difference at the same row/frame-rate budget. The worst-direction hard radius is approximately $0.50k_N$ for SLIM versus $0.25k_N$ for the frame-rate-matched control, while SLIM approaches the native cutoff along favorable directions. The FI-defined bandwidth follows the same trend but is smaller because it additionally reflects OTF attenuation and multi-view weighting (Fig. 5(c)). Table 3 summarizes the key quantitative metrics derived from this simulation.

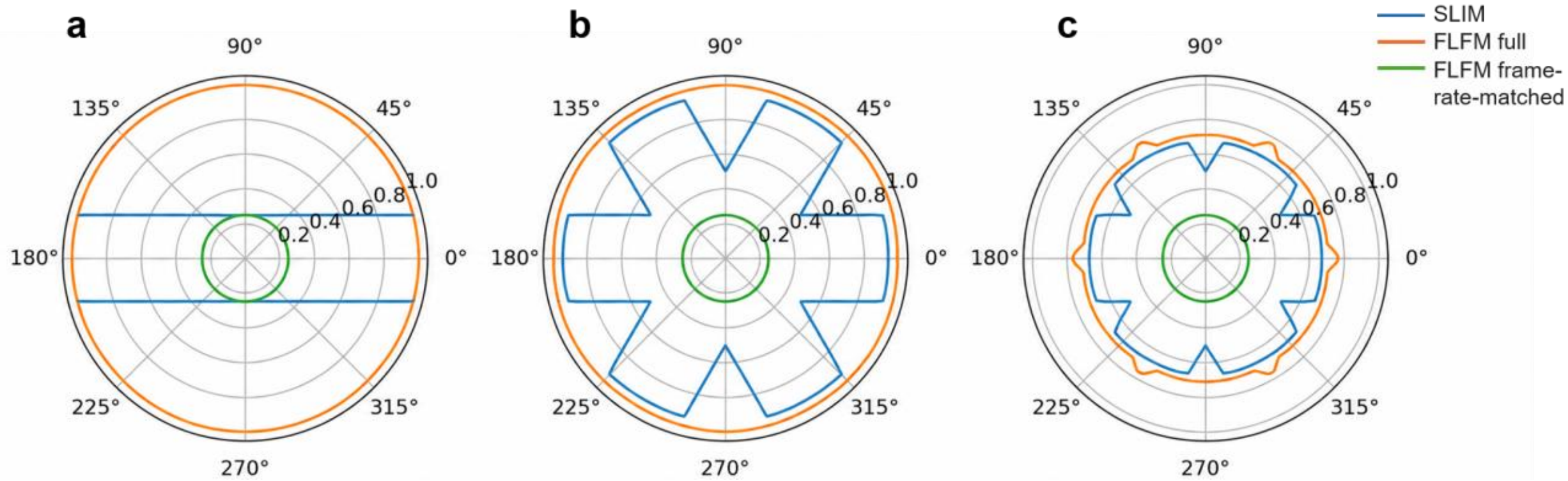

**Figure 5. Native support, projected hard support, and FI-defined bandwidth under the camera-bandwidth comparison.** (a) Single-view in-slice support: full FLFM is a unit disk, frame-rate-matched FLFM is a disk of radius s, and SLIM retains native disparity-axis extent while compressing the transverse axis. (b) Radial boundary of the binary three-view projected support. The weak modulation of full FLFM is a projection effect; the stronger SLIM structure arises from genuine anisotropic sampling combined with complementary view rotations. (c) FI-defined radial bandwidth at the 0.01 threshold, showing that SLIM retains substantially larger mean and worst-direction information bandwidth than frame-rate-matched FLFM.

Figure 6 summarizes these complementary metrics and shows that, although full FLFM retains the largest unconstrained 3D support, SLIM preserves the full axial-frequency reach while providing substantially greater spatial-frequency coverage and integrated FI per camera bandwidth than frame-rate-matched FLFM. These trends persist over a broad range of compression factors and view tilts (Supplementary Note 2; Supplementary Fig. S1). The advantage increases with stronger compression and converges toward unity as $s \rightarrow 1$, while the $1/s$ axial-frequency scaling remains robust to the Fourier-slice tilt within the present model.

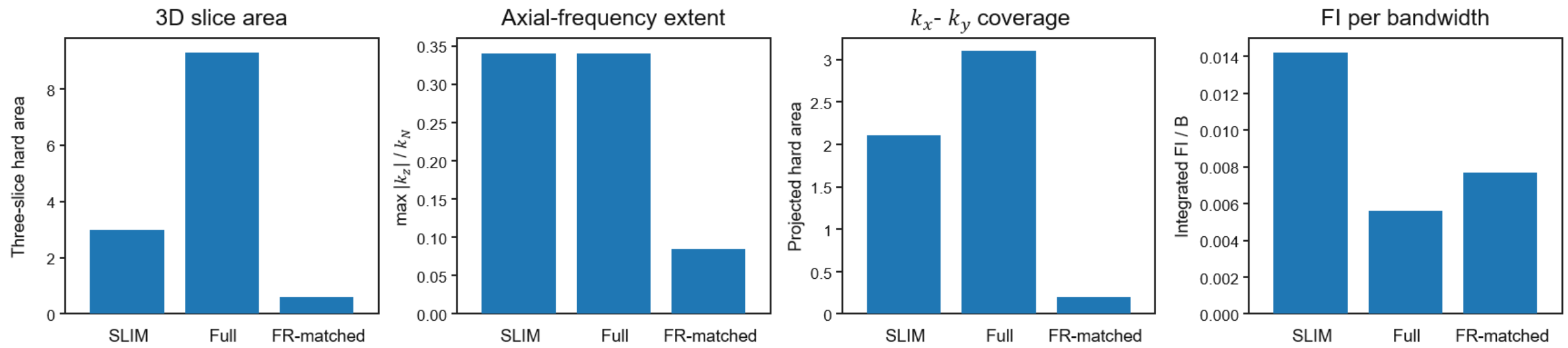

**Figure 6. Complementary dense-scene metrics under the row-limited camera model**: 3D slice-support area, hard axial-frequency extent, projected $k_x - k_y$ hard-support area, and integrated Fourier-mode FI/B. Full FLFM maximizes unconstrained support and total information, whereas SLIM preserves the full-FLFM axial reach while using one quarter as many active rows. Relative to frame-rate(FR)-matched FLFM, SLIM provides broader 3D support and higher integrated information per row-readout cost.

Taken together, the sparse- and dense-scene analyses identify the same mechanism from two complementary parameterizations. Disparity-aware squeezing does not create information beyond that available to full FLFM; instead, it preferentially retains the parameter directions and spatial frequencies most useful for 3D inference while reducing the detector resource that sets frame time. The discussion below separates this task dependence from the detector-architecture dependence and clarifies the role of full FLFM as the unconstrained reference.

## 5. Discussion

The results support a unified view of detector-efficient optical encoding. Encoder performance is governed by two distinct choices: the scene determines which parameter directions or spatial-frequency modes must be distinguished, whereas the camera architecture determines the cost of acquiring those distinctions. The sparse and dense analyses should therefore not be reduced to a single universal ranking. Rather, they show how a common anisotropic encoding redistributes information under different inverse problems and how the value of that redistribution depends on the detector resource that limits acquisition rate.

### 5.1 Scene model determines the relevant Fisher information

Classical FI is local to a specified parameterization and does not, by itself, impose a sparsity prior. Here the scene regime enters through the unknowns used to define the FIM. An isolated emitter is described by a small set of coordinates; a crowded sparse scene requires a joint multi-emitter position FIM whose off-diagonal blocks quantify source ambiguity; and a dense continuous object is more naturally described by spatial-frequency coefficients. The resulting metrics are related but not interchangeable: point-localization FI quantifies coordinate precision, whereas Fourier-mode FI quantifies the observability of object frequencies.

This distinction is consistent with information-theoretic studies of multiplexed imaging showing that the preferred encoder depends on object sparsity and detection noise[8]. Sparse scenes can tolerate stronger multiplexing because relatively few encoded PSFs overlap. As occupancy increases, source-identity ambiguity becomes an important loss mechanism, appearing here as stronger FIM correlations and smaller eigenvalues. In the dense limit, overlap is intrinsic and the relevant question shifts from identifying individual sources to preserving and conditioning spatial-frequency or voxel modes. The scene model must therefore define what constitutes useful information before detector efficiency is assessed.

A more explicit statistical treatment of sparsity could extend the framework by averaging the FIM over scene ensembles with prescribed occupancy statistics[8], using restricted-support FIMs[12], or introducing Bayesian/Van Trees priors[13,14]. Such approaches would provide a continuous bridge between the isolated-emitter, crowded-source, and dense-volume limits considered here.

### 5.2 Detector architecture governs the benefit of anisotropic squeezing

Once the estimation task is fixed, the detector architecture determines how efficiently an encoder acquires the relevant information. At equal active-row count, frame-rate-matched FLFM reduces both in-plane sampling dimensions by the squeeze factor. SLIM instead retains the full column dimension and compresses only the row direction after rotating each view so that its disparity-sensitive coordinate lies along the preserved axis. The three preserved directions are distributed over the 0°, 120°, and 240° views, so the retained column samples provide complementary directional information rather than repeated copies of the same measurement. In a column-parallel detector, these samples do not carry the same frame-time penalty as additional rows; FI per total pixel is therefore not the relevant efficiency metric for the architecture considered here.

The Fourier-slice model gives this detector-allocation argument a direct 3D interpretation. On a tilted slice, the disparity-aligned coordinate $k_u$ is geometrically coupled to $k_z$. SLIM retains $|k_u|$ up to the native cutoff while frame-rate-matched FLFM limits both in-plane coordinates to $sk_N$. The hard axial-frequency extent therefore differs by approximately 1/s; at $s$ = 0.25, SLIM has a 4× advantage over the isotropic control while matching the full-FLFM axial reach. The complementary view rotations also improve lateral coverage: for three preserved axes, the guaranteed worst-direction hard-support radius is approximately $\min(1,2s)k_N$ for SLIM versus $sk_N$ for frame-rate-matched FLFM, giving a 2× advantage when $s \leq 0.5$.

The same resource matching changes the interpretation of integrated FI. Under an equal-pixel comparison, an isotropically demagnified control can retain a relatively broad low-frequency disk because its linear demagnification scales as $\sqrt{s}$. Under equal row bandwidth, the control must instead demagnify both axes by $s$, discarding substantially more transverse and disparity-linked bandwidth. At $s$ = 0.25, SLIM consequently provides 1.85× higher integrated Fourier-mode FI/B than frame-rate-matched FLFM. This scalar efficiency gain complements rather than replaces the support metrics: integrated FI emphasizes information magnitude and redundancy, whereas hard support, directional isotropy, and worst-mode bandwidth specify where that information remains accessible.

### 5.3 Total information versus information rate

Full FLFM is the appropriate unconstrained reference because it retains the complete native in-plane disk for every view and therefore maximizes high-frequency reach and transverse redundancy within the present model. SLIM should not be interpreted as containing more information per frame than full FLFM. Its advantage emerges after the detector readout resource is included: at $s$ = 0.25, full FLFM uses four times as many active rows and therefore has one quarter of the nominal frame rate in the assumed row-limited architecture. After subpixel-phase averaging, SLIM provides 4.00× higher axial FI/B and 3.24× higher D-optimal position FI/B than full FLFM in the isolated-emitter model; in the dense model, its integrated Fourier-mode FI/B is 2.54× higher. These ratios quantify information-rate efficiency under the specified detector architecture, not greater unconstrained optical information content.

### 5.4 Scope and limitations

The analysis intentionally isolates detector sampling and readout architecture, and the conclusions should be interpreted within that scope. The sparse model uses a pixel-integrated Gaussian PSF, whereas the dense model uses ideal projection/Fourier slices with a circular in-plane OTF. A real segmented-pupil fluorescence microscope has finite-thickness 3D OTF support determined by diffraction, defocus, aberrations, pupil geometry, and calibration[15-17]. A system-specific extension should therefore begin with measured or wave-optically calculated 3D PSFs for the native views, apply the corresponding SLIM and FLFM detector mappings, and recompute the Poisson FIM from the sampled measurements.

The detector and scene models are also simplified. The point-source calculation includes Poisson shot noise and a small uniform background but omits read noise, gain noise, pixel-response nonuniformity, finite well depth, calibration uncertainty, and other camera-specific effects that may shift the optimum squeeze factor. The isolated-emitter analysis averages over a finite 4 × 4 detector-phase ensemble; denser phase sampling or experimental averaging would further suppress residual discretization dependence. Most importantly, FI/B assumes that frame time is governed primarily by active-row count and that additional columns are digitized in parallel without an equivalent throughput penalty. Real detectors may instead be limited by ROI width, ADC sharing, interface bandwidth, packet-transfer overhead, or other bottlenecks. For a specific camera, active-row count should therefore be replaced by the measured frame-time or throughput model. Likewise, the dense Fourier analysis uses weak perturbations around a locally uniform background; application-specific predictions will require representative scene ensembles.

## 6. Conclusion

A fair information-theoretic comparison of detector-efficient imaging systems must account for both the inverse problem and the architecture of the detector used to acquire the measurements. The scene determines which Fisher-information representation is relevant, whereas the detector architecture determines the acquisition resource against which that information should be normalized. In the column-parallel, row-limited camera model considered here, active-row count provides an appropriate proxy for readout cost. Under this constraint,

SLIM and frame-rate-matched FLFM use the same number of active rows and therefore have the same nominal frame rate, while SLIM exploits the parallel detector-column dimension to preserve sampling along the disparity-sensitive direction.

For the baseline squeeze factor $s = 0.25$, this anisotropic allocation preserves essentially the full-FLFM axial localization information while providing 2.40× higher axial FI/B and 1.89× higher 3D D-optimal FI/B than frame-rate-matched FLFM. For dense Fourier modes, SLIM provides 1.85× higher integrated FI/B, 5.05× greater 3D slice-support area, 4× greater hard axial-frequency extent, and approximately 11× greater projected lateral hard-support area at the same nominal frame rate. The additional robustness analysis shows that these advantages are not specific to this operating point but instead reflect a systematic principle of detector-resource allocation.

More generally, the framework developed here is not specific to SLIM, FLFM, or even row-limited cameras. Its central requirement is to combine a task-appropriate Fisher-information model with an architecture-specific measure of detector acquisition cost. For a conventional column-parallel camera, this cost may be approximated by active-row count or measured frame time; for other detector architectures, the framework can be adapted by replacing both the measurement likelihood and the acquisition-cost normalization with architecture-specific forms. For example, an event-based camera could be evaluated using event-generation and event-transfer throughput, whereas a time-delay-integration camera could be normalized according to line-readout rate, charge-transfer stages, or the corresponding effective acquisition time. More generally still, detector throughput, digitization rate, data-transfer capacity, or directly measured acquisition time can be used to define $C_{det}$ whenever they better describe the hardware bottleneck.

The broader design principle is therefore that optical encoding should be co-designed with both scene statistics and detector readout architecture. When some detector degrees of freedom are less costly than others, an encoder can preferentially allocate measurements to those dimensions while preserving the parameter directions or spatial frequencies most important to the inference task. Fisher information normalized by the appropriate detector resource provides a general framework for quantifying this trade-off and for comparing optical encoders across different imaging tasks and camera architectures.

**Acknowledgements**

Funding support from National Institutes of Health (R01HL16531, RF1NS128488, R35GM128761); U.S. Department of Energy (DE-SC0025928).